\documentclass{ws-procs9x6}

\begin{document}

\title{Phenomenological analysis of the CLAS data on double charged pion
photo and electro- production off proton.}

\author{V. ~I. Mokeev, V. ~D. Burkert and  L. Elouadrhiri}

\address{Jefferson Lab, \\
12000 Jefferson Ave., Newport News, VA 23606, USA \\ }

\author{A. ~A. Boluchevsky, G. ~V. Fedotov, E ~L. Isupov, B. ~S. Ishkhanov and 
  N. ~V. Shvedunov}

\address{Skobeltsyn Nuclear Physics Institute at Moscow State University, \\ 
119899 Vorobevy gory, Moscow, Russia \\}  

\maketitle

\abstracts{First comprehensive data on the evolution of nucleon resonance photocouplings 
with photon virtuality $Q^{2}$ are presented for excited proton states in the
mass range from 1.4 to 2.0 GeV. $N^{*}$ photocouplings were determined in 
phenomenological analysis of CLAS data on 2$\pi$ photo and 
electroproduction within the framework of the JLAB-MSU phenomenological model.}

\section{Introduction}

Comprehensive studies of nucleon resonances were carried out in phenomenological analysis of 
CLAS data on double charged pion production by
real and virtual photons \cite{Ri03,Be04}. The analysis was performed within the framework
of the 2005 version of  JLAB-MSU phenomenological model in the following referred to
as JM05.

\section{2$\pi$ model. From 2003 to 2005 version.}

Electromagnetic production of two pions from proton is sensitive to contributions 
from both low-lying
and high-lying $N^{*}$ states. This exclusive channel offers an opportunity to determine
electromagnetic transition form factors from the nucleon ground state to many 
excited states.  Moreover, it is also
a promising channel in the search for so-called "missing" $N^{*}$ states. These 
states are predicted 
from symmetry principles of the symmetric constituent quark models. 
Many of these states are expected to couple strongly to N$\pi$$\pi$ final state.
However, $N^{*}$ decays into the 
2$\pi$ channel contribute only a fraction of the total 2$\pi$ production; a large
part is due to non-resonant production mechanisms. 
In such conditions, the development of
reaction models, which provide reasonable background treatment and $N^{*}$/background 
separation, become necessary for the evaluation of $N^{*}$ parameters.

We have developed a phenomenological model, that incorporates particular 
meson-baryon mechanisms
based on their manifestations in observables$:$ as enhancements in invariant
mass distributions, sharp forward/backward slopes in angular distributions.
The p$\pi$$\pi$ final state offers a
variety of single differential cross sections for analysis. Any particular 
meson-baryon mechanism has distinctive reflections in different
cross sections.
Therefore, a combined analysis of 
an entire set of single
differential cross sections makes it possible to establish the relevant
meson-baryon mechanisms from the data fit. 
Our model is currently limited to the N$\pi^{-}$$\pi^{+}$ final state and incorporates
 particular
meson-baryon mechanisms needed to describe $\pi^{+}$p, $\pi^{+}$$\pi^{-}$,
$\pi^{-}$p invariant masses and
$\pi^{-}$ angular distribution. These cross sections were analyzed in the
hadronic mass range from 1.41 to 1.89 GeV and in four $Q^{2}$ bins  
 centered at 0., 0.65, 0.95, 1.30 $GeV^{2}$. The overall $Q^{2}$-coverage ranges 
from 0. to 1.5 $GeV^{2}$.

In the 2003 version (JM03) \cite{Mo01,Bu03}
double charged pion 
production was described by the 
superposition of quasi-two-body channels with the formation and
subsequent decay of unstable particles in the intermediate states:
\begin{eqnarray}
         \gamma p \rightarrow \pi^{-} \Delta^{++} \rightarrow
\pi^{-}  \pi^{+} p, \\
     \gamma p \rightarrow \pi^{+} \Delta^{0} \rightarrow
\pi^{+} \pi^{-} p,  \\
     \gamma p \rightarrow \rho^{0} p \rightarrow
\pi^{+} \pi^{-} p. \\
     \gamma p \rightarrow  \pi^{+} D^0_{13}(1520) \rightarrow
\pi^{+} \pi^{-} p.
\end{eqnarray}
Remaining residual mechanisms  were 
parametrized as 3-body phase 
space with the amplitude fitted to the data. 
This amplitude was a function of photon 
virtuality $Q^2$ and
invariant mass of the final hadronic system W only. In this 
approach we were able to reproduce
the main features of integrated cross sections as well as
$\pi^{+}\pi^{-}$, $\pi^{+}p$ invariant masses and  $\pi^{-}$ 
angular distributions in CLAS electroproduction data \cite{Ri03,db}.

The production amplitudes for the first three quasi-two-body 
mechanisms (1-3) were treated as sums
of $N^{*}$ excitations in the $s$-channel and non-resonant mechanisms
described in Refs \cite{Mo01,Bu03}.
The quasi-two-body  mechanism (4) was
entirely non-resonant \cite{Mo04}.
In reactions (1-3) all well established 4 star resonances
with observed decays to the two pion
final states were included as well as the 3-star states
$D_{13}(1700)$, $P_{11}(1710)$, $P_{33}(1600)$, and $P_{33}(1920)$. For the $P_{33}(1600)$
a 1.64 GeV mass was obtained in our fit. This value
is in agreement with the results of recent analyses of 
$\pi N$ scattering experiments.
$N^{*}$ electromagnetic transition  
form factors were
fitted to the data.
Hadronic couplings for $N^*\rightarrow \pi \Delta$
and $\rho p$ decays were taken from the analyses
of experiments with hadronic probes, 
except for $P_{33}(1600)$, $P_{13}(1720)$, the candidate $3/2^{+}(1720)$,
$D_{13}(1700)$, $P_{33}(1920)$, $F_{35}(1905)$, and  $F_{37}(1950)$
states. Poorly known hadronic decay parameters
for these states were fitted to the data.

Analysis of CLAS 2$\pi$ electroproduction data within the framework of this
approach revealed the structure around 1.7 GeV, which can not be explained
by the contributions from conventional $N^{*}$ only \cite{Ri03}. We found
two possible ways to describe CLAS data around W=1.7 GeV: a) assuming
drastically different $\pi$$\Delta$ and $\rho$p hadronic couplings for the
$P_{13}(1720)$ state with respect to the established couplings, or b) keeping hadronic decay
parameters for all $N^{*}$$'$s inside established uncertainties, a new baryon state 
with quantum numbers
$3/2^{+}(1720)$ 
is needed to describe CLAS data around W=1.7 GeV.

\begin{figure}[ht]
\centerline{\epsfxsize=4.1in\epsfbox{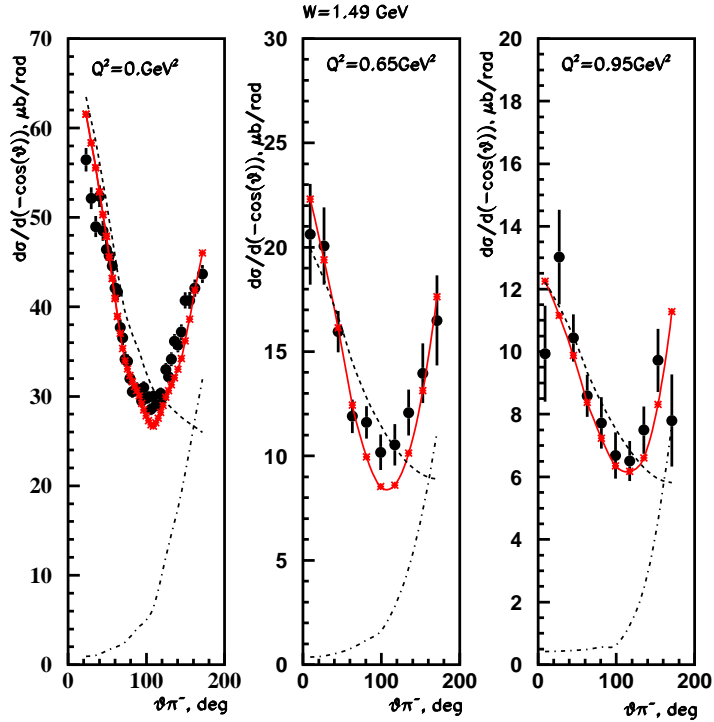}}   
\caption{ $\pi^{-}$ CM angular distributions at W=1.49 GeV and at three photon
virtualities. CLAS data [3-5] are shown in comparison
with the JM05 results$:$ solid lines represent full 
calculations$;$
the contributions from 2$\pi$ direct production mechanisms are shown by
dashed-dotted lines. The full calculations within the framework of the  
JM03 are shown by dashed lines.
\label{ang149}}
\end{figure}

\begin{figure}[ht]
\centerline{\epsfxsize=4.1in\epsfbox{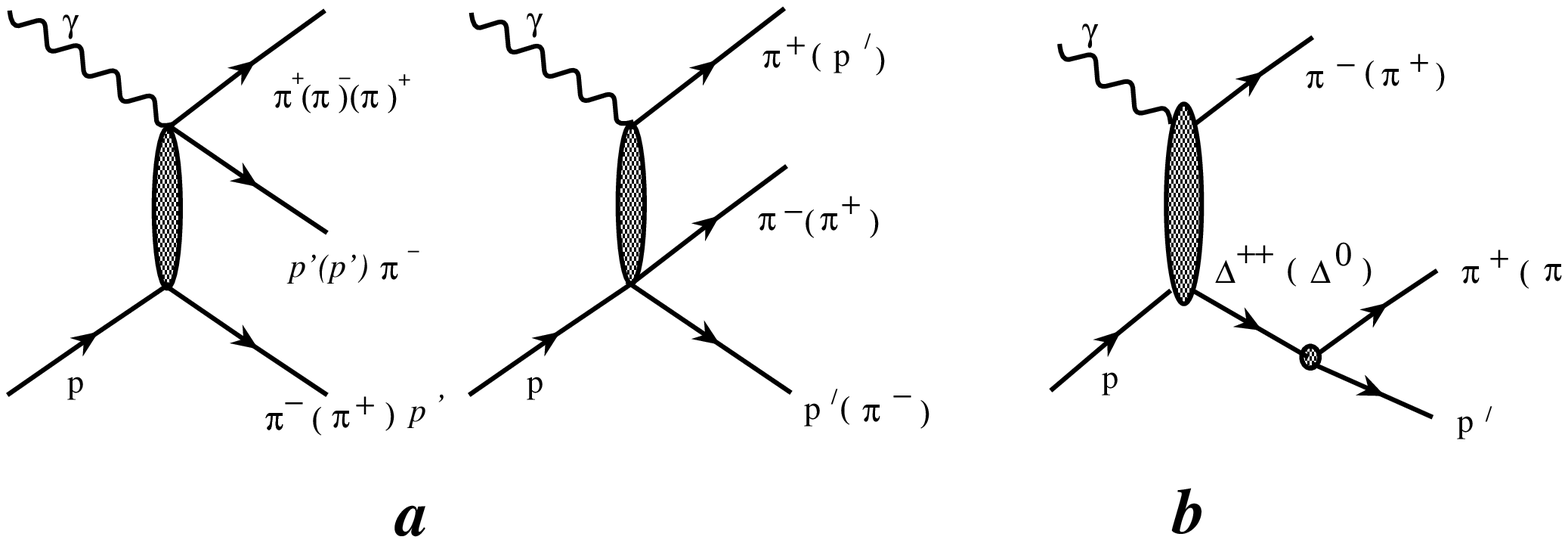}}   
\caption{Direct 2$\pi$ production mechanisms (a). Complementary contact term 
in $\pi$$\Delta$ production (b). 
\label{diag}}
\end{figure}

The analysis of preliminary CLAS data on 2$\pi$ photoproduction \cite{Be04} within the
framework of JM03
revealed shortcomings in the description of $\pi^{-}$ angular distributions at backward 
hemisphere (Fig.~\ref{ang149}). Similar incompatibilities were also
seen for electroproduction data \cite{Ri03,db}. They are related to
the parametrization of remaining mechanisms as 3-body phase, 
 which  is incompatible with the steep increase of the measured
$\pi^{-}$  angular distributions at backward angles. In JM05, 
the 3-body phase space description
was replaced 
by the set of exchange
terms shown in Fig.~\ref{diag}a. This allowed much improved 
description
of the $\pi^-$ angular distributions (solid lines in Fig.~\ref{ang149} )
in the entire $Q^{2}$ range
covered by the CLAS data. Parametrization of these exchange amplitudes 
is described in \cite{Az05}.

\begin{figure}[ht]
\centerline{\epsfxsize=4.1in\epsfbox{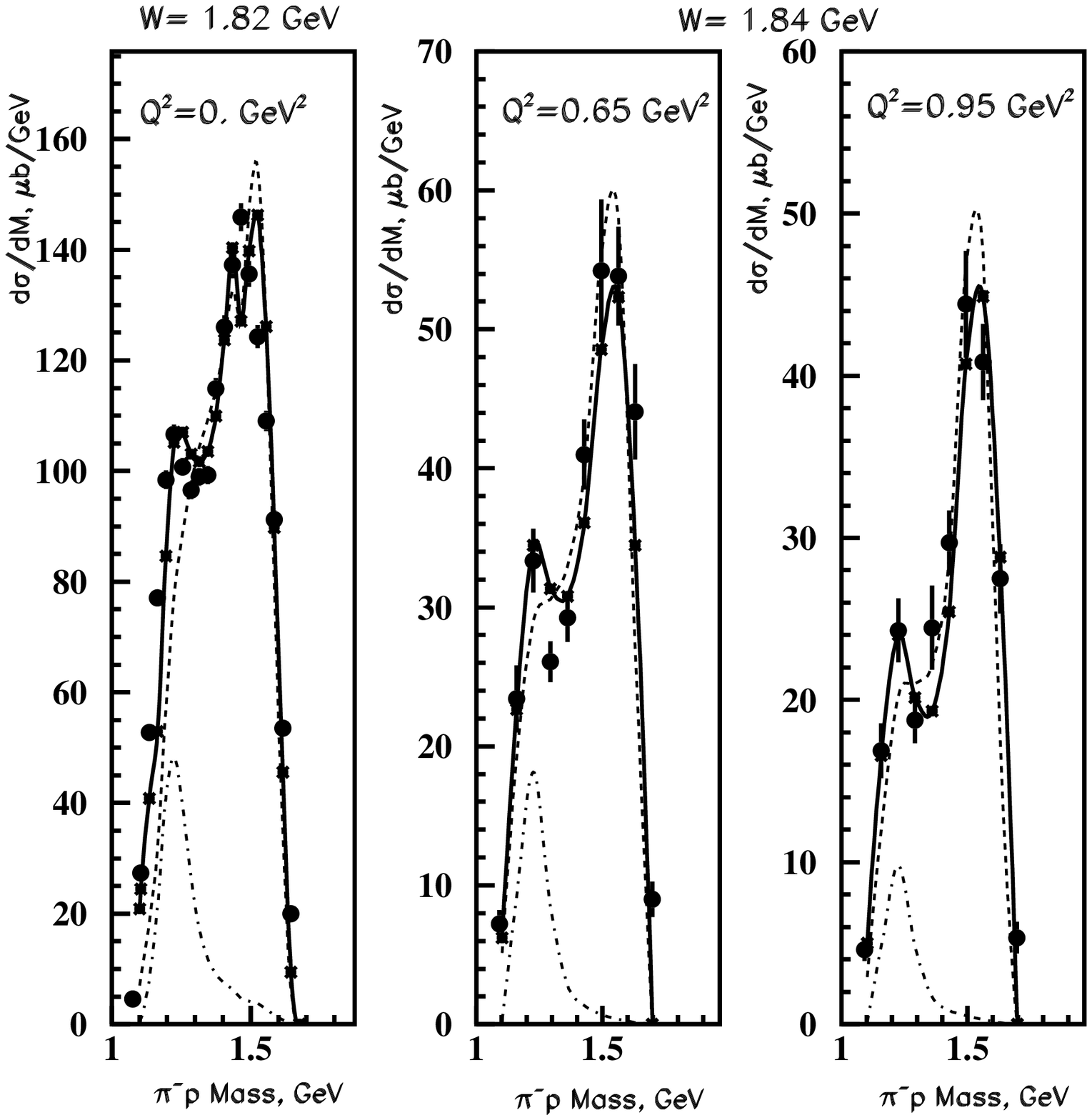}}   
\caption{Evidence for complementary contact term contribution to the $\pi$$\Delta$
isobar channels. Full calculations in JM05 are shown by solid lines,
while evaluations within the framework of JM03 are shown by
dashed lines. $\pi^{+}$$\Delta^{0}$ channel contributions estimated in
JM05 model are shown by dotted-dashed lines. CLAS 2$\pi$ photo and
electroproduction data are from [3-5]  
\label{tensor}}
\end{figure}

Such modifications for remaining mechanisms allow us to reproduce  
CLAS data on 2$\pi$ photo- and electroproduction
\cite{Ri03,Be04,db}  at W$\leq$1.7 GeV reasonably well. Above 1.7 GeV the 
measured cross sections around 
$\Delta^{0}$ mass in the $\pi^{-}$p mass distributions exceed the model
cross sections (Fig
~\ref{tensor}).
To reproduce the strength of the $\Delta^{0}$ peaks (solid lines in Fig.~\ref{tensor} ) 
as well as to improve description of the
$\pi^{+}$p mass distributions  we implemented an
additional contact term for isobar channels (1)-(2), shown on Fig.~\ref{diag}c. 
Parametrization of these mechanisms is presented in \cite{Mo05c}.


\begin{figure}[ht]
\centerline{\epsfxsize=4.1in\epsfbox{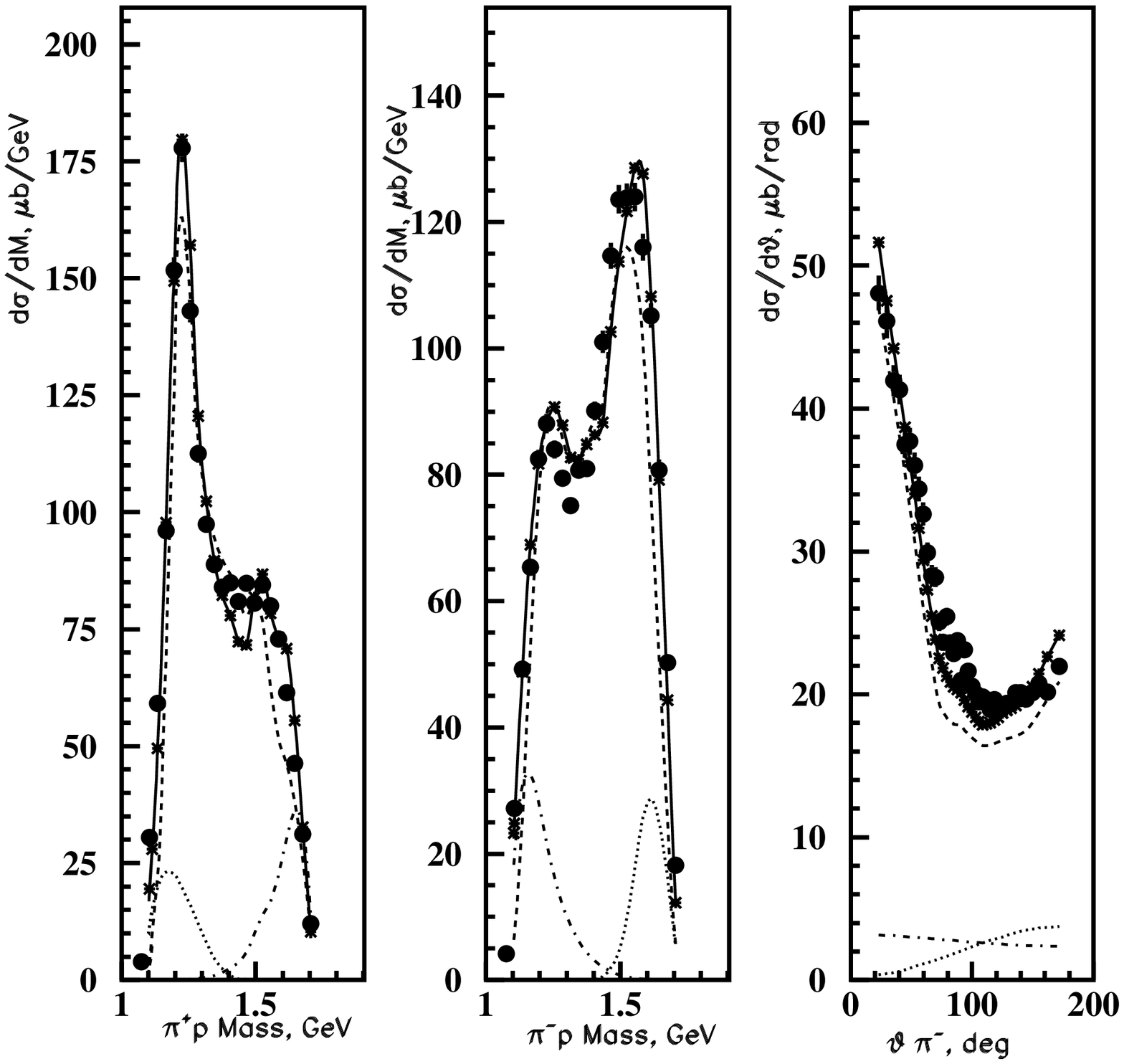}}   
\caption{Manifestation of $\pi^{+}$$F_{15}^{0}(1685)$ and
$\pi^{-}$$P_{33}^{++}(1600)$ isobar channels. The photoproduction CLAS data 
at W=1.86 GeV 
are compared to the JM05: full calculations (solid lines)$;$
 isobar channels (5),(6) 
are absorbed in 3-body phase space (dashed lines). The
contributions from channels (5),(6) are
shown by dotted and dotted-dashed lines respectively.  
\label{isonew}}
\end{figure}

After these improvements we still have shortcomings in the description of 
$\pi^{+}$p, $\pi^{-}$p mass and $\pi^{-}$-angular distributions at W above 
1.8 GeV (dashed lines in Fig.~\ref{isonew}). Lack of strength in the calculated 
$\pi^{+}$p, $\pi^{-}$p mass distributions are centered at the masses of 
$P_{33}(1660)$ (1.64 GeV fitted mass) and $F_{15}(1685)$ resonances,
respectively. The gap in angular distributions may be filled by t-channel exchange
mechanisms for $\pi^{+}$$F_{15}^{0}(1685)$ intermediate state. The observed 
discrepancies thus
indicate contributions from the isobar channels:
\begin{eqnarray}
         \gamma p \rightarrow \pi^{+} F^0_{15}(1685) \rightarrow
\pi^{+}  \pi^{-} p, \\
     \gamma p \rightarrow \pi^{-} P^{++}_{33}(1600) \rightarrow
\pi^{-} \pi^{+} p.  
\end{eqnarray}  
Implementation of isobar channels (5)-(6) with amplitudes as outlined in
\cite{Mo05c} allow us to reproduce the data at W$\geq$ 1.8 GeV reasonably well 
(solid lines in
Fig.~\ref{isonew}).

After these modifications, we succeeded in describing all
available CLAS data on unpolarized observables in 2$\pi$ photo 
and electroproduction. These results are presented 
in \cite{Mo05c}. We found no need for 
remaining mechanisms of unknown dynamics. 
Therefore, the quality of the CLAS data allow us to establish all significant mechanisms
in 2$\pi$ production, implementing particular meson-baryon 
diagrams as determined from the data fit. 

The credibility of our description of non-resonant mechanisms and the separation of
resonant and non-resonant contributions was tested in a combined analysis of CLAS data on 1$\pi$ and 2$\pi$
electroproduction \cite{Az05}. We found
a common set of  $N^{*}$ photocouplings, which allowed us to reproduce all observables
measured in these two exclusive channels combined. Since 1$\pi$ and
2$\pi$ channels represent two major contributors in $N^{*}$ excitation region 
with considerably different background, their successful fit
offers compelling evidence for credible background description
and N*/background separation achieved in JM05.

\section{$N^{*}$ analysis in 2$\pi$ photo and electroproduction.}

\begin{figure}[ht]
\centerline{\epsfxsize=4.1in\epsfbox{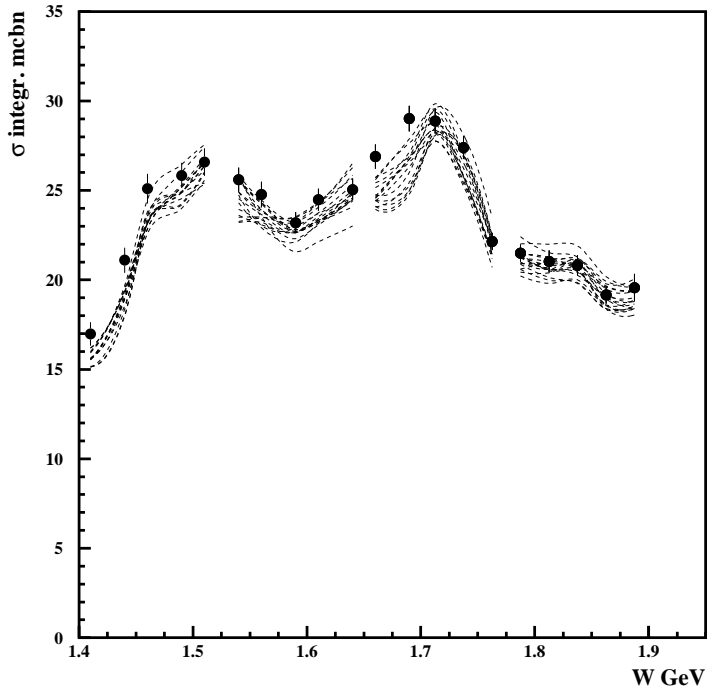}}   
\caption{Total 2$\pi$ electroproduction cross sections calculated
within the framework of JM05 with $N^{*}$ parameters fitted to the
CLAS data [3-5] at photon virtuality 0.65 $GeV^{2}$.
\label{total}}
\end{figure}

We fit all available CLAS data on 2$\pi$ photo and electroproduction at
W$\leq$.1.9 GeV and $Q^{2}$ from 0 to 1.5 $GeV^{2}$ within the framework of
JM05. $N^{*}$ photocouplings were sampled 
according to the
normal distribution around the values, obtained 
in the JM03 \cite{Mo05cc}. The photocouplings were varied
within 0.3 $\sigma$ from theirs starting values. Poorly 
known masses and
hadronic couplings were also 
fluctuated inside the uncertainties established in experiments with hadronic probes. 
Adjustable
parameters of non-resonant mechanisms were varied within 0.2 $\sigma$.
For
each trial set of model parameters we calculated all kind of single
differential cross sections in all available W and $Q^{2}$
bins.
From comparison between calculated and measured single differential 
cross sections
$\chi^{2}$/d.p. were estimated. We isolated a bunch of calculated cross sections
inside the data uncertainties, applying restriction
$\chi^{2}$$\leq$$\chi^{2}_{th}$, where $\chi^{2}_{th}$ is a predetermined maximal
allowed value.  
Integrated 2$\pi$ cross sections in comparison with selected 
calculated cross sections are shown in Fig.~\ref{total}. A  reasonable description
of all cross sections was achieved.

$N^{*}$ photocouplings  for selected cross sections were averaged and 
mean values were treated as extracted from the data fit, while
dispersions were assigned to photocoupling uncertainties. In this way we obtained the
photocouplings for the states: $P_{11}(1440)$, $D_{13}(1520)$, $S_{31}(1620)$,
$S_{11}(1650)$, $P_{33}(1600)$,  $F_{15}(1680)$,
$D_{13}(1700)$, $D_{33}(1700)$, candidate $3/2^{+}(1720)$, $P_{13}(1720)$,
$F_{35}(1905)$, $P_{33}(1920)$, and $F_{37}(1950)$ \cite{Mo05c}. 

\begin{figure}[ht]
\centerline{\epsfxsize=4.1in\epsfbox{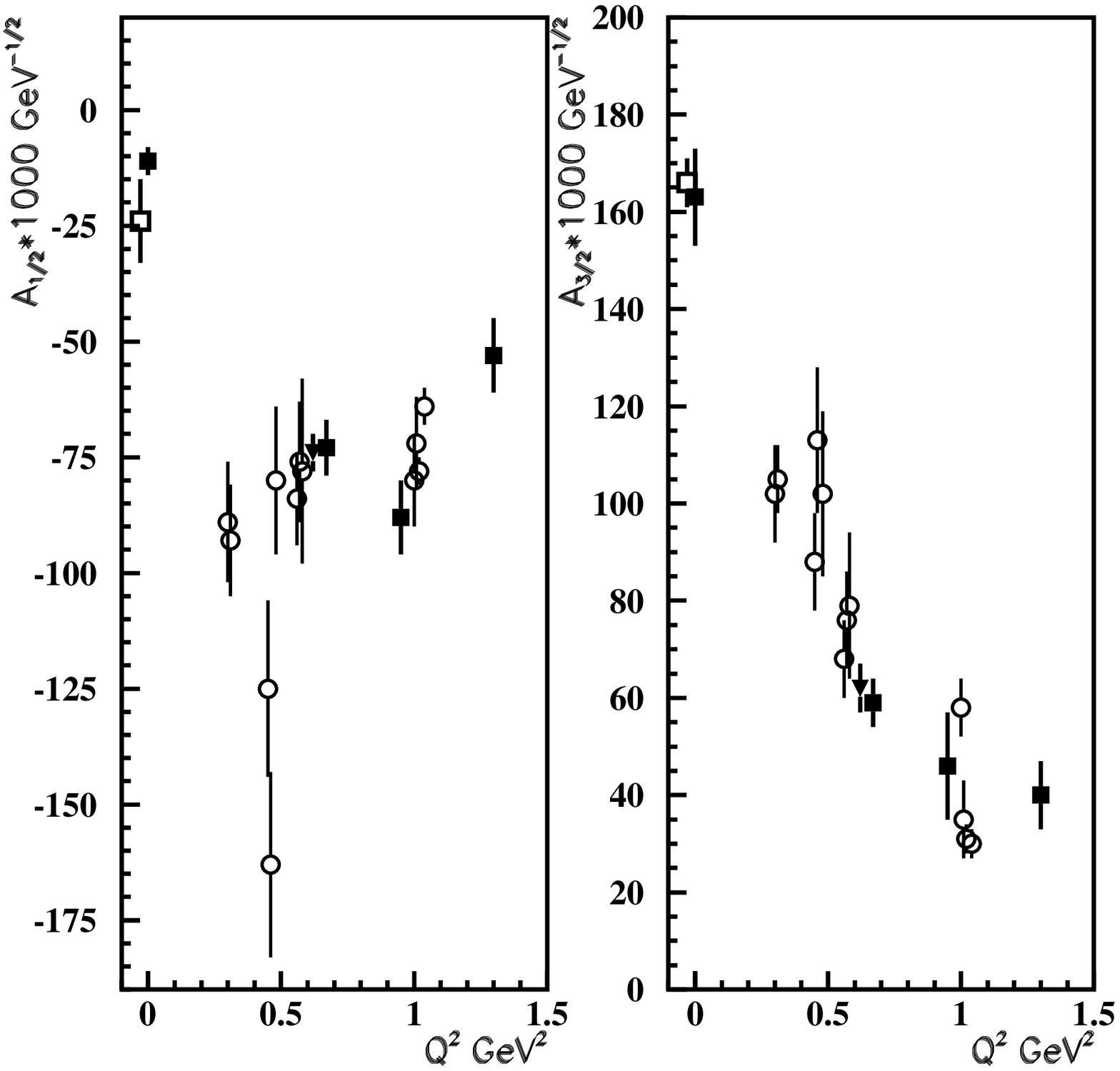}}   
\caption{$D_{13}(1520)$ photocouplings extracted from the analysis of CLAS 2$\pi$ data
[3-5] (filled squares) in comparison with world data [10] (open circles) and 
results of analysis of the CLAS 1$\pi$ and 2$\pi$ data combined [7] (filled
triangles).  
\label{d131520}}
\end{figure}

In Fig.~\ref{d131520} we present the
photocouplings for the well studied $D_{13}(1520)$ state in comparison with
previously available data.
Reasonable overlap between our results and previous ones support the 
reliability of our
procedure for the extraction of  $N^{*}$ photocoupling from 2$\pi$ data fit. 
For the first time, we determine the
electrocouplings for high lying $N^{*}$, which preferably decay with 2$\pi$
emission: $D_{13}(1700)$, $D_{33}(1700)$, candidate
$3/2^{+}(1720)$, $P_{13}(1720)$, $F_{35}(1905)$, 
$P_{33}(1920)$, and $F_{37}(1950)$.
\begin{figure}[ht]
\centerline{\epsfxsize=4.1in\epsfbox{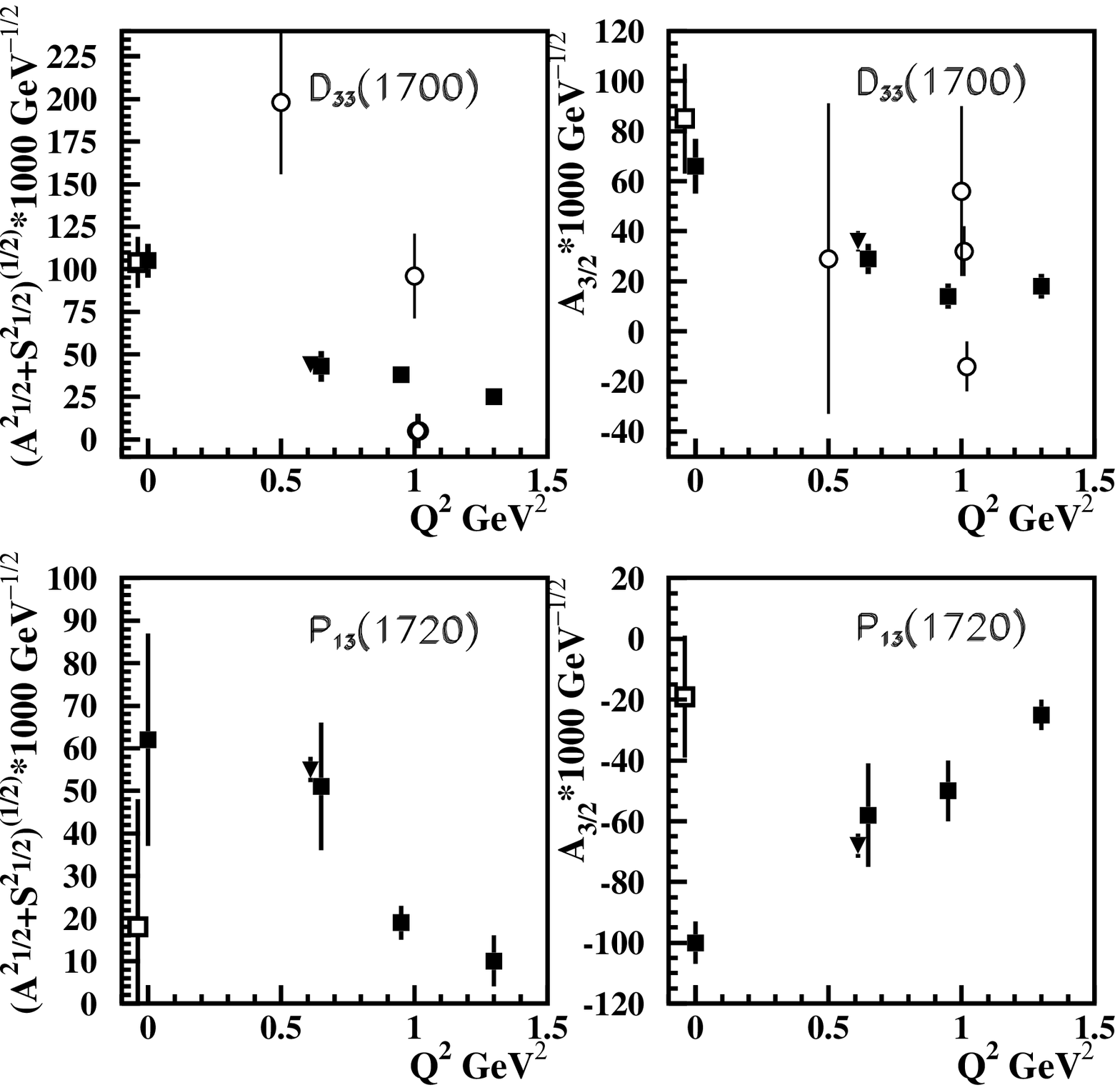}}   
\caption{$D_{33}(1700)$, $P_{13}(1720)$ photocouplings extracted from analysis of 
CLAS 2$\pi$ data in comparison with previous results. Symbols are the same as
in the Fig. 6.  
\label{d331700p131720}}
\end{figure} 
In Fig.~\ref{d331700p131720}  we present photo- and electrocouplings for the 
$D_{33}(1700)$ and  
$P_{13}(1720)$ states extracted from the CLAS 2$\pi$ data, as
well as couplings obtained in previous studies of 1$\pi$ production. 
The analysis of the CLAS 2$\pi$ data for the first time provides 
accurate information on
electrocouplings for high lying nucleon excitations.  

\section{Conclusions.}

A phenomenological model for the description of 2$\pi$ production from protons in the nucleon
resonance
region was developed with most complete accounting for all
relevant mechanisms. The reliability of the background treatment and $N^{*}$/background
separation was confirmed by the reasonable description obtained for of all unpolarized
observables in this exclusive channel, as well as in the combined analysis of
1$\pi$ and 2$\pi$ production. Electromagnetic transition form factors were extracted
at photon virtualities $Q^{2}$ $\leq$ 1.5 $GeV^{2}$   
for $N^{*}$ states in the mass range from 1.4 to 2.0 GeV. For the first time transition form
factors were obtained for
many high lying proton states with major 
2$\pi$ decay.

\end{document}